\documentclass[acus]{JAC2003}

\usepackage{multirow}
\usepackage{cite}
\usepackage{graphicx}
\usepackage{booktabs}
\usepackage{epstopdf}

\begin{document}
\title{MITIGATION OF COLLECTIVE EFFECTS BY OPTICS OPTIMISATION}

\author{Y.~Papaphilippou,  F.~Antoniou and H.~Bartosik, CERN, Geneva, Switzerland}

\maketitle

\begin{abstract}
This paper covers recent progress in the design of optics solutions to minimize collective effects 
such as beam instabilities, intra-beam scattering or space charge in hadron and lepton rings. 
The necessary steps are reviewed for designing the optics of high-intensity and high-brightness 
synchrotrons but also ultra-low emittance lepton storage rings, whose performance is strongly 
dominated by collective effects. Particular emphasis is given to proposed and existing designs 
illustrated by simulations and beam measurements.
\end{abstract}

\section{INTRODUCTION}

The physics quantities parametrising the performance of a large variety of hadron and lepton rings,
as, for example, the average beam power of synchrotron based proton drivers, the luminosity of colliders, 
the brightness of their associated injectors or the brilliance of X-ray storage rings,
are proportional to the beam intensity or to its ratio with the beam dimensions. 
The modern tendency is to push the performance frontiers towards extreme conditions, i.e.
the highest beam intensity contained within ultra-low beam volumes,
under which the collective behaviour of the beam becomes predominant. It is thus of paramount
importance to take measures in order to alleviate collective effects 
in the early phase of the design, which usually begins with the optimisation of the linear optics.

This is by far not an easy task, as already there is a large amount of optics conditions based 
on single-particle constraints to be satisfied, including non-linear dynamics. 
Thereby, the parameter space becomes entangled and difficult to control and optimise. One
should rely on analytical and numerical methods for obtaining a global parameterisation, 
including collective effects.

In the case of rings in operation, dealing with collective effects usually implicates mitigation
techniques based on the use of multi-pole magnets~\cite{chao} or higher harmonic RF cavities~\cite{theo} 
for providing Landau damping,  dedicated feedback systems~\cite{hofle} or the reduction of the 
beam interaction with its environment through careful vacuum and low-impedance component design~\cite{paul}.
Changing the linear optics, without major upgrade involving radical modifications of the machine configuration,
 is an unconventional approach, since it is subject to the constraints of 
the existing magnet and powering systems. It can be even more challenging 
because of its interplay with the already optimised operation of critical systems, such as beam transfer elements or RF. 
On the other hand, if a viable solution is found, it can be a very cost effective way to overcome existing 
intensity or brightness limits. In addition, it may be used to relax  
tolerances associated with the above-mentioned mitigation measures for reducing collective effects.

This paper is organised as follows:
after describing the basic linear optics parameters which affect collective effects,  optics design strategies 
are reviewed, underlining specific examples  of high-power or high-brightness synchrotrons and low emittance damping rings,
studied in recent years, mainly at CERN. 
Of particular interest is the application of these approaches to operating rings with illustrations from the 
direct impact of the optics modifications to machine performance.

\section{Impact of optics parameters on collective effects}

In this section,  three fundamental  quantities that
affect collective effects are described, following the logical route
of an optics study: starting from the most basic one, the beam energy, 
passing to the most fundamental, the transverse beam sizes and ending
with the phase slip factor, which will be shown to be 
most intimately connected to the collective beam behaviour.

\subsection{Beam energy}
The beam energy is one among the basic parameters that have to be specified even
before starting the optics design of a ring. Although, strictly speaking, 
it cannot be considered as an optics constraint, it is indirectly related
 through the integrated magnet strengths and the size of the lattice cells, i.e.
the ring circumference. At the same
 time, in the absence of strong synchrotron radiation damping, the transverse emittance is inversely proportional to the energy, thus reducing the
 physical beam size. Almost all collective effects become less pronounced with increasing beam energy, with the notable exception of
 the electron cloud instability thresholds~\cite{giovprl}.
 Hence, for hadron rings, it is natural to target always the highest possible energy although this heavily depends on the 
 users' physics needs, the reach of the pre-injectors and finally on cost. In the case of beams dominated by
 synchrotron radiation damping, e.g. for ultra-low emittance e$^+$/e$^-$ rings, the quadratic dependence of the horizontal equilibrium emittance 
 to the energy puts an additional restriction to this increase, and a careful optimisation has to be performed, in order to meet the specific design targets
 and reaching high brightness.

\subsection{Betatron functions}
Transverse beam sizes are also playing an important role to the collective beam behaviour, especially in the case of self-induced
 fields. For example, the space-charge tune-shift~\cite{laslett} and IBS growth rates~\cite{ibs} are inversely proportional
 to their product raised to a certain power.  For high-intensity/power rings, there is usually no specific preference on the size of
 transverse emittances and the trend is to produce them large enough, for limiting the aforementioned effects. When the performance 
 target is high brightness, 
 which corresponds to small (transverse) emittances, the optics is the only "knob" for increasing beam sizes. For hadron rings, the FODO cells are 
 well suited for this, due to the alternating behaviour of the optics functions. In particular, weaker focusing
 can maximise not only betatron beam sizes but also their dispersive part (through the contribution of momentum spread and dispersion),
 within the limits set by the machine aperture. In the case of e$^+$/e$^-$ rings targeting low emittances, 
 doublet-like cells are usually employed for minimising horizontal beam sizes. On the other hand, the vertical beta functions can be increased, 
 especially along the bending magnets, where the horizontal ones are small. Although this strategy is valid for space-charge or IBS, 
 beam current thresholds of instabilities such as transverse mode coupling  or coupled bunch, present an opposite dependence and
 call for a reduction of the average (vertical) beta functions. 
 Finally, the enhancement of betatron functions at the location of non-linear magnetic elements can provide additional tune-shift with amplitude
 for Landau damping in case of need~\cite{fartoukh}.

\subsection{Slippage factor}
The slippage (or phase slip) factor $\eta$ is defined as the rate of change of the revolution frequency 
with the momentum deviation. At leading order, 
it is a function of the relativistic  $\gamma$ factor (i.e. the energy) and the momentum compaction factor $\alpha_p$:
\begin{equation}
\label{Eq:slippagefactor}
\eta=\alpha_p-\frac{1}{\gamma^2}\;.
\end{equation}
The momentum compaction factor is the rate of change of the circumference $C$ with the momentum spread and, again at leading
order, it is given by
\begin{equation}
\label{Eq:momcomp}
\alpha_p = \frac{1}{C} \oint \frac{D_x(s)}{\rho(s)}ds\;,
\end{equation}
which depends clearly on the variation of the horizontal dispersion function along the bending magnets. The phase slip factor
 unites transverse and longitudinal particle motion. In fact, the synchrotron frequency
or the bunch length are proportional to $\eta^{1/2}$, which means that increasing the slippage factor makes synchrotron
motion faster, with an equivalent increase of the bunch length.
 
 The phase slip factor vanishes when 
$\gamma = \alpha_p^{-1/2} \equiv \gamma_t$ and the corresponding energy is named transition energy. It is widely known, 
since the commissioning
of the first synchrotrons, 
that crossing transition can cause various harmful effects with respect to the collective behaviour of the beam~\cite{transcros1}, as the longitudinal motion basically freezes, at this point. 
Although several transition crossing schemes have been proposed and operated reliably in synchrotrons like 
the CERN PS for more than 40 years (see ~\cite{transcros2} and references therein),
the call for beams with higher intensity (or power) resulted in the consideration of ring designs which avoid transition, either by injecting above ($\eta>0$), or
always remaining below ($\eta<0$) transition energy. The former case is almost always true for electron/positron rings above a few hundred MeV (unless $\alpha_p<0$).
For hadron rings, it requires the combination of high energy (i.e. large circumference) and a large momentum compaction, which is translated to larger
dispersion excursions and, generally speaking, weaker focusing, thereby resulting in larger beam sizes~\cite{Ng}. For remaining below transition, the operating energy range 
has to be kept narrow and a positive momentum compaction factor should be low, which points towards stronger focusing and smaller beam sizes.
The case of negative momentum compaction (NMC)~\cite{FMC} is indeed very interesting because the beam remains always below transition independent 
of energy. Again, as in the case of the rings remaining above transition, the need to excite dispersion oscillations for getting an overall negative
dispersion integral on the bends, results in larger beam sizes.

The above discussion is even more interesting when combined with the dependence of  intensity thresholds for most transverse and longitudinal
instabilities to the absolute value of the slippage factor~\cite{chao}. A large slip factor provides additional spread in the synchrotron tunes, thereby increasing Landau damping.   
Although the particular characteristics of each machine may direct to different optics optimisation routes, the
above mentioned simple considerations trace some generic guidelines for reducing collective effects, i.e. increase of the slippage factor (in absolute) combined
with increased beam sizes can be achieved simultaneously above transition, or below transition and negative momentum compaction. Remaining
below transition has the additional benefit of enabling the damping of the lowest head-tail instability  modes with negative chromaticity~\cite{chao}, as the natural one,
hence avoiding the use of strong sextupoles which excite resonances and induce beam losses.

\section{High-power synchrotrons}

\begin{figure}[h]
    \centering
    \includegraphics[trim = 5 5 0 23, clip,width=0.8\columnwidth]{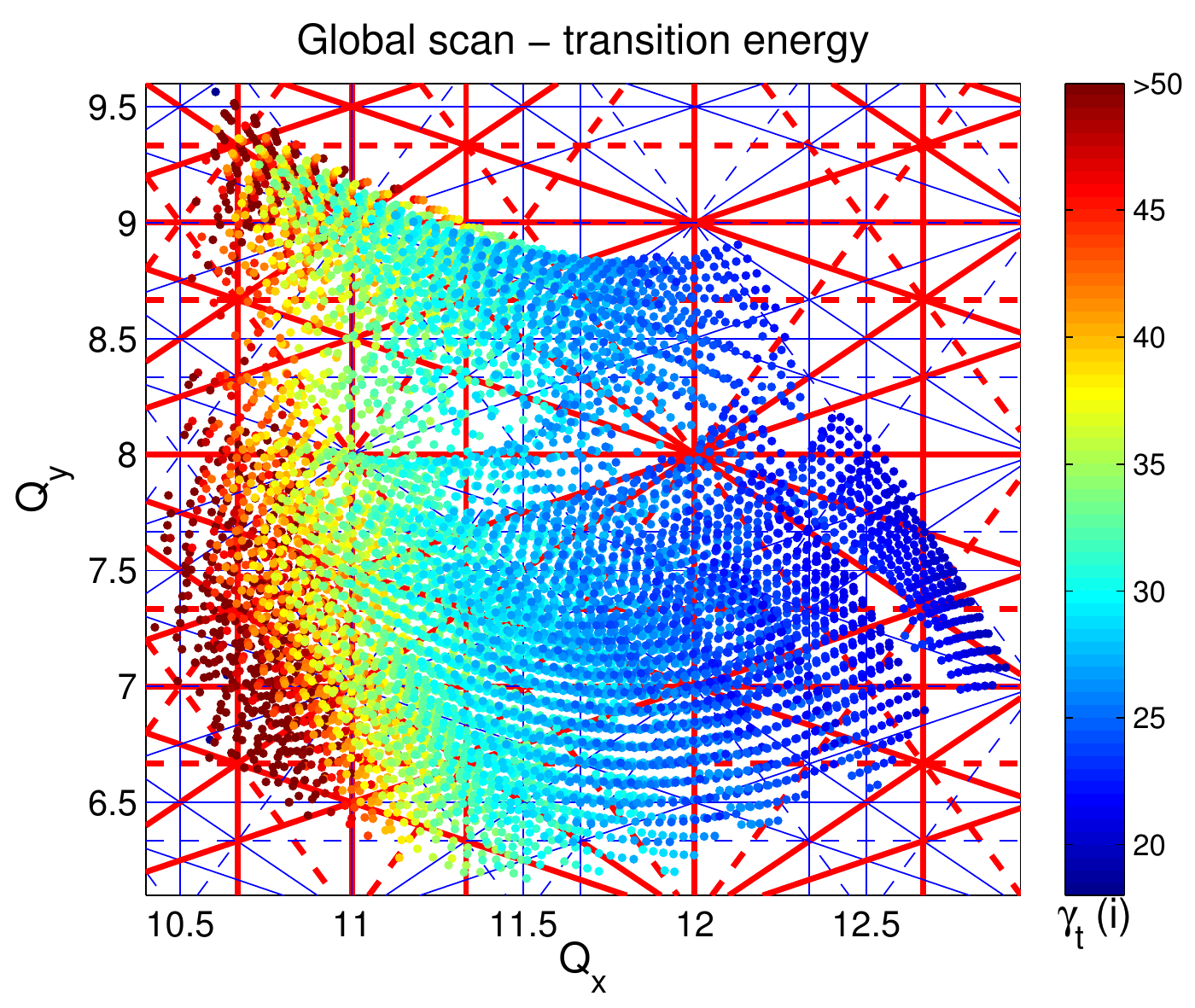}
    \includegraphics[trim = 5 5 0 23, clip,width=0.8\columnwidth]{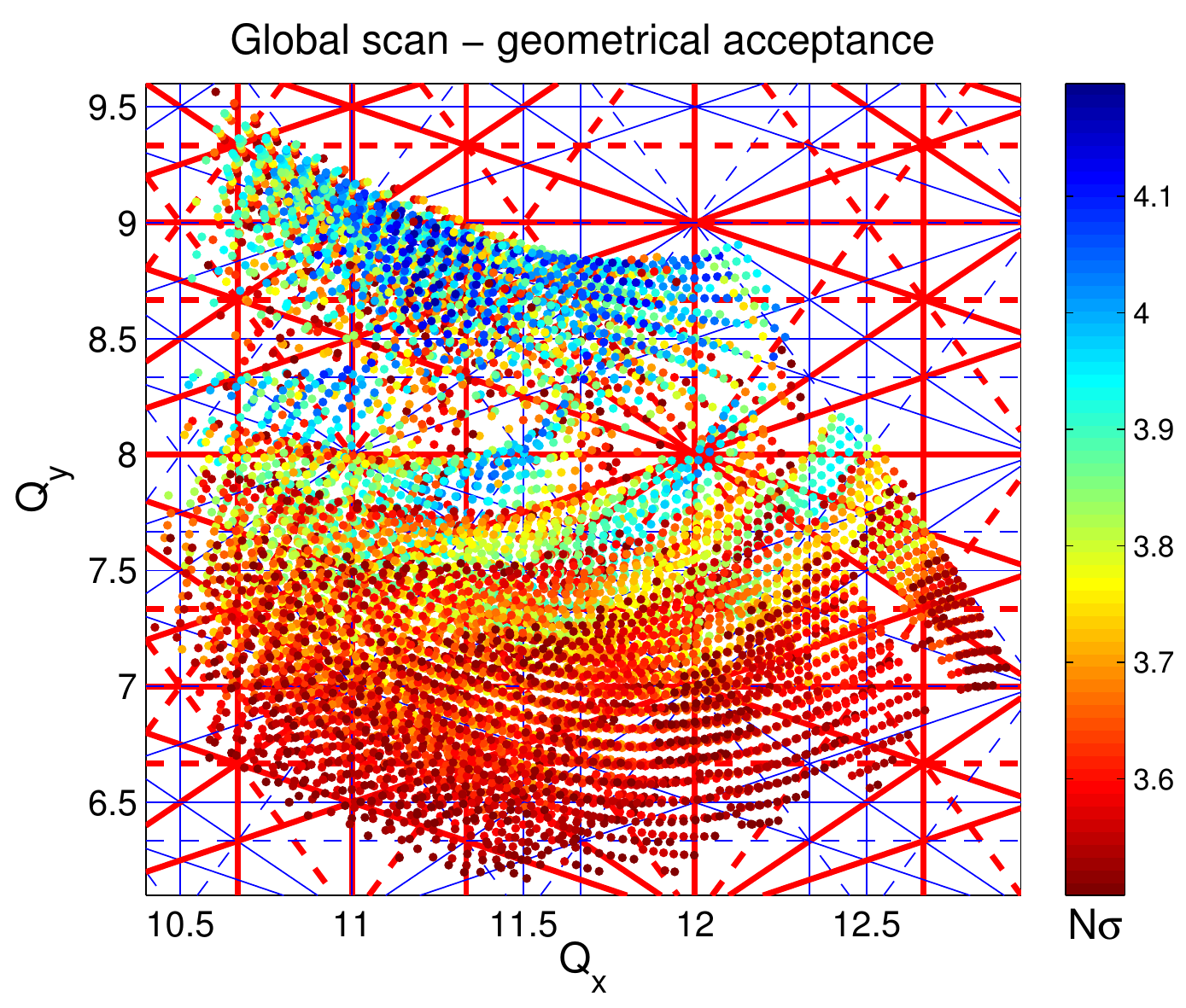}
    \caption{Transition energy $\gamma_t$ (top)  and geometrical acceptance in units of beam sizes N$_\sigma$ (bottom), for a global scan of optics solutions in the tune diagram (showing resonances up to $3^{rd}$ order), with blue corresponding to lower $\gamma_t$ or larger acceptance~\cite{bib:PS2GLASS}.}
    \label{FIG:Tunability}
\end{figure}

Recent optics design of high-intensity and/or high-power rings such as the J-Parc main ring~\cite{J-Parc}, the PS2~\cite{PS2}, or the High-Power PS~\cite{HPPS} 
are based on NMC arc cells, for avoiding transition and reducing losses. These are sequences of modified FODO cells 
with an increased number of quadrupole families (up to four) for inducing negative dispersion, leading to an overall ``imaginary" $\gamma_t$~\cite{FMC}. In that case, the 
absolute value of the slippage factor could be increased for raising instability thresholds but also because a fast synchrotron frequency would be beneficial
for longitudinal beam manipulation~\cite{steve}. A complete picture of the achievable tuning range of a ring such as the PS2 can be obtained  by the 
Global Analysis of all Stable Solutions (GLASS), a numerical method pioneered in low emittance rings~\cite{ALSGLASS}, where all possible quadrupole configurations
(within some gradient limits) providing stable solutions are obtained, together with the optics parameters associated to them. In the top part of Fig.~\ref{FIG:Tunability}, the imaginary transition $\gamma_t$ is presented for all stable solutions in the tune diagram, along with resonance lines up to 3rd order. 
Low imaginary values of $\gamma_t$ (i.e. large absolute values of the momentum compaction), indicated in blue are obtained for higher horizontal tunes, 
where there is large flexibility for the vertical tunes.
In the bottom part of the figure, the geometrical acceptance is computed for the most demanding beam parameters with respect to emittance. The red colour corresponds to small
 acceptance (above a limit of 3.5~$\sigma$), which means larger beam sizes. The trend shows that the larger sizes (red colour) are obtained for lower vertical tunes. This type of global analysis including linear and non-linear dynamics constraints was used for choosing the working point during the conceptual design of the PS2 ring, and locate it at ($Q_x$,$Q_y$)=(11.81,6.71), with $\gamma_t = 25.3i$~\cite{bib:PS2GLASS}.
 
\section{Low Emittance Rings}

The present trend of ultra-low emittance rings is to target the highest beam intensities
within the smallest dimensions, at least in the transverse plane. The additional complication
in the case of damping rings (DRs) for linear colliders is that they aim to produce low longitudinal emittances,
as well. The output beam dimensions are largely dominated by IBS and even space-charge effects become important,
especially in the vertical plane. A careful optimisation of the optics parameters  is crucial  for reducing these effects 
and obtaining a solid conceptual design~\cite{fanouriaphd}. 

\subsection{Mitigating collective effects in the CLIC DRs}

\begin{figure}[t]
\centering
    \includegraphics[trim = 10 0 0 24, clip, width=0.49\columnwidth]{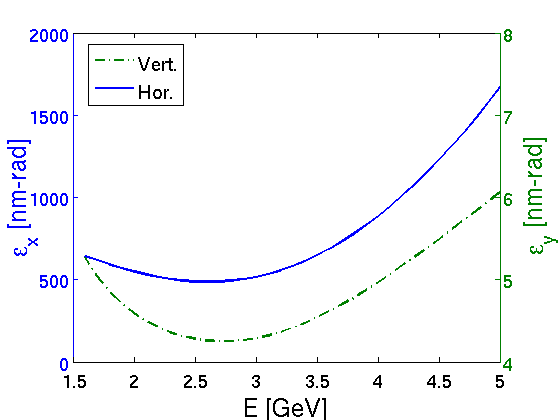}
    \includegraphics[trim = 10 0 0 24, clip, width=0.49\columnwidth]{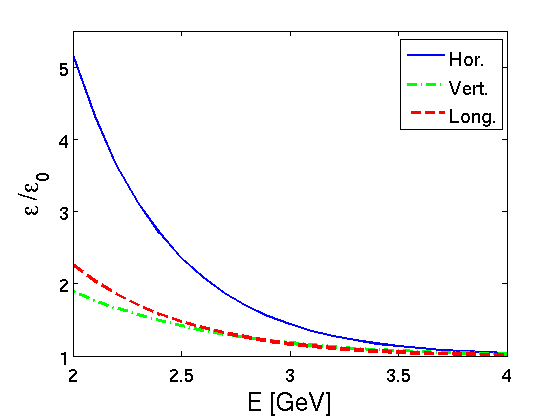}
    \caption{Steady-state emittances (left) and their blow-up (right) due to IBS, as a function of the energy~\cite{fanouriaphd}.}
\label{FIG:IBSenergy}
\end{figure}

Due to the fact that not only the IBS growth rates but also the equilibrium emittances vary with energy,  
it is important to find their interdependence, when the IBS effect is included~\cite{scan}. 
Evaluated through a modified version of the Piwinski  method~\cite{CIMP}, and for constant
longitudinal emittance, the dependence of the steady state transverse emittances of the CLIC DRs on
the energy is plotted in
Fig.~\ref{FIG:IBSenergy} (left). A broad minimum is observed around 2.6~GeV for both horizontal (blue) and vertical planes (green).
The IBS effect becomes weaker with the increase of energy, as shown in Fig.~\ref{FIG:IBSenergy} (right), where
the emittance blow-up for all beam dimensions is presented.
Although higher energies may be desirable for reducing further collective effects, 
the output emittance is increased above the target value, due to the domination of quantum excitation.
In this respect, it was decided to increase the CLIC DR energy to 2.86~GeV, already reducing the IBS impact 
by a factor of two, as compared to earlier designs at 2.42~GeV~\cite{scan}.

\begin{figure}[h]
\begin{center}
    \includegraphics[trim = 0 0 0 0, clip, width=0.24\textwidth]{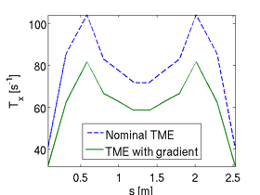}
    \includegraphics[trim = 0 0 0 0, clip, width=0.24\textwidth]{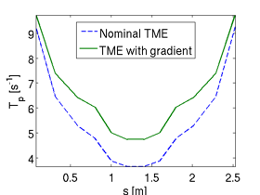}
\caption{The horizontal (left) and longitudinal (right) IBS growth rate evolution for a standard TME cell (blue dashed) and a TME cell with
a dipole with gradient (green).}
\label{FIG:TME_IBS}
\end{center}
\end{figure}

In modern low emittance rings, theoretical minimum emittance (TME) arc cells or multi bend
achromats are employed. In order to reach minimum emittance, the horizontal beam
optics is quite constrained, whereas the vertical one is free, but also completely determined
by the two quadrupole families of the cell. It turns out that the vertical beta function
reaches a minimum at the same location as the horizontal, which is the worst case for IBS. 
A way to reverse this tendency, is to use a combined function dipole
with a low defocusing gradient. Although this gradient does not provide a significant effect
to the emittance reduction, it reverses the behaviour of the vertical beta function
at the middle of the dipole, maximizing the vertical beam
size at that location, and thus reducing IBS growth rates~\cite{newcliclat}. This is shown in Fig.~\ref{FIG:TME_IBS},
where the horizontal (left) and longitudinal (right) IBS growth rate evolution are presented 
for a standard TME cell (blue dashed lines) and compared to the corresponding ones when the
dipole includes a vertical focusing gradient (green curves). A reduction of the horizontal
growth rate of almost a factor of two can be achieved in the shown example corresponding to the
CLIC damping rings arc cell, by allowing a smaller increase to 
the longitudinal IBS growth rate. 

\begin{figure}[t]
\centering
    \includegraphics[trim = 10 0 5 5, clip, width=0.49\columnwidth]{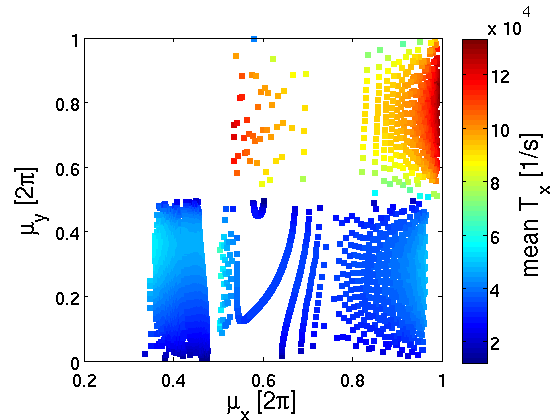}
    \includegraphics[trim = 10 0 5 5, clip, width=0.49\columnwidth]{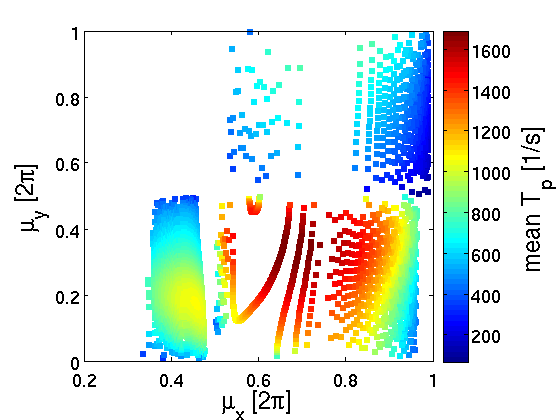}\\
    \includegraphics[trim = 10 0 5 5, clip, width=0.49\columnwidth]{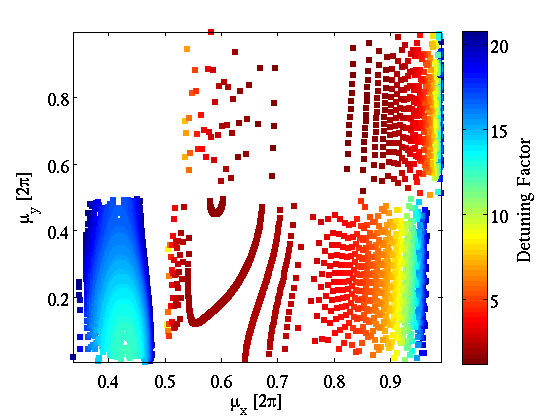}
    \includegraphics[trim = 10 0 5 5, clip, width=0.49\columnwidth]{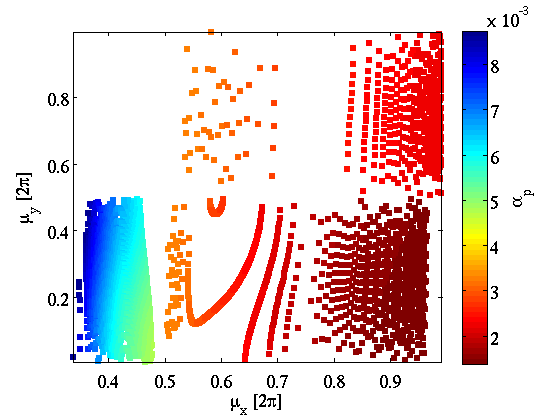}\\
    \includegraphics[trim = 10 0 5 5, clip, width=0.49\columnwidth]{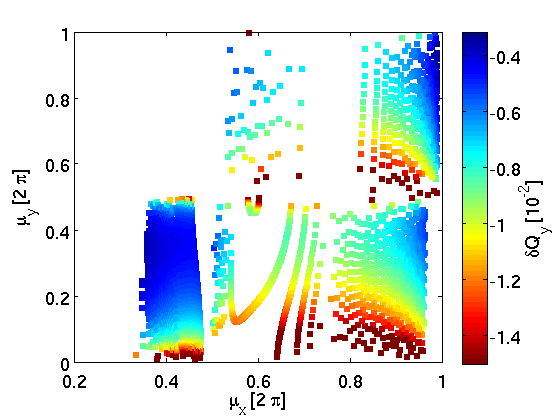}
    \includegraphics[trim = 10 0 5 5, clip, width=0.49\columnwidth]{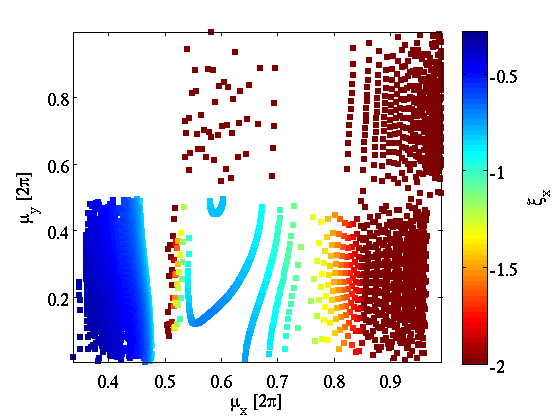}
    \caption{
Analytical parameterization of the TME cell phase advances with the IBS horizontal (top, left) and longitudinal (top, right) growth
rates, the detuning factor (middle, left), the momentum compaction factor
(middle, right),  the horizontal chromaticity (bottom, left) and the Laslett tune shift (bottom, right)~\cite{fanouriaphd}.}
\label{FIG:muxmuy}
\vspace{-10pt}
\end{figure}

A crucial step in the optimisation of the TME cell with respect to its impact on collective effects
is the analytical derivation of the two quadrupole focal lengths,
in thin lens approximation, depending only 
on the horizontal optics functions at the centre of the dipole and the drift
space lengths~\cite{fanouriaphd,TMEanalytical}. Using this representation, the dependence of various parameters  on the cell
phase advances in the case of the CLIC DRs are presented in Fig.~\ref{FIG:muxmuy}, including the average IBS growth rates 
(top), the detuning from the minimum emittance (middle, left) the momentum compaction factor (middle, right),
the vertical space-charge tune-shift (bottom, left) and the horizontal chromaticity (bottom, right). This parameterisation
permitted to find the best compromise for the phase advances (between 0.4 and 0.5) where the IBS growth rates, 
the horizontal and vertical chromaticities and the Laslett tune shift are minimized,
while the momentum compaction factor is maximized. These low phase advances correspond
to emittances that deviate from the absolute minimum by a factor of around 15, as shown in Fig.~\ref{FIG:muxmuy} (middle left).
Even at this large detuning factor, the TMEs are preferable for their compactness, in particular for a ring in which
radiation damping is dominated by wigglers. The use of variable bends with gradient
 in the TME cell was also studied using a similar approach, further reducing the IBS growth rates~\cite{PhysRevAccelBeams.22.091601}.

\begin{figure}[t]
\begin{center}
    \includegraphics[trim = 0 0 0 0, clip, width=0.24\textwidth]{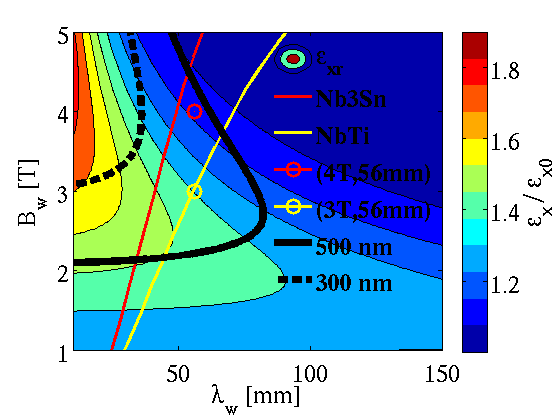}
    \includegraphics[trim = 20 30 20 20, clip, width=0.23\textwidth]{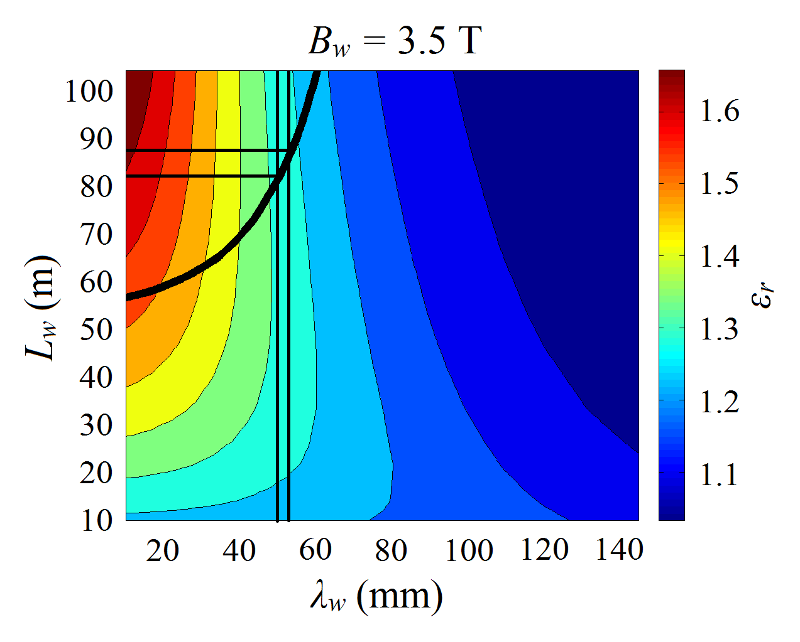}
    \caption{Dependence of the steady state emittance ratio with the equilibrium
emittance as a function of the wiggler peak field and period (left)~\cite{fanouriaphd,wiggler} and as a function of the total wiggler length and period
for a 3.5~T peak field (right)~\cite{Fajardo:2016nnq}.}
\label{FIG:wiggler}
\end{center}
\end{figure}

A similar study was performed in order to find the optimal wiggler field and wavelength, while minimising the IBS effect~\cite{fanouriaphd,wiggler, Fajardo:2016nnq,PhysRevAccelBeams.22.091601}.  In Fig.~\ref{FIG:wiggler} (left),  the steady state emittance ratio with the equilibrium
emittance as a function of the wiggler peak field and period is presented~\cite{fanouriaphd,wiggler}. The limits
for the two superconducting technologies are shown in yellow (NbTi) and red (Nb3Sn) and the
300~nm and 500~nm target steady state emittance contours in black. Based on these studies, 
the highest field within the limit of technology would be desirable, but a moderate wavelength
is necessary for reducing IBS. At the same time, as shown in Fig.~\ref{FIG:wiggler} (right), by raising the field and using Nb3Sn wire technology, 
the reduction of the ring circumference can be also achieved, with beneficial impact to all type of collective effects, including to a potential impedance reduction~\cite{Fajardo:2016nnq,PhysRevAccelBeams.22.091601}. These specifications were used for the super-conducting
wiggler prototype and short model developed for the CLIC DRs~\cite{wiggler,Fajardo:2016nnq}.

\section{High-Brightness Synchrotrons}

Hadron collider injectors need to achieve the highest brightness with the smallest possible losses. 
A typical example is the CERN SPS whose performance limitations and their mitigations for LHC beams are the subject of a study group~\cite{elena},
 in view of reaching the required beam parameters for the high luminosity LHC (HL-LHC). The upgrade of the main 200~MHz RF system will solve beam loading 
 issues for reaching higher intensities, but a variety of single and multi-bunch instabilities remain to be confronted. The transverse mode coupling instability (TMCI) in the vertical
plane and e-cloud instability (ECI) for 25~ns beams are the most prominent transverse problems, especially for HL-LHC intensities. Longitudinal 
  instabilities necessitate the use of a higher harmonic 800~MHz RF system as Landau cavity and the application of controlled longitudinal
  emittance blow-up throughout the ramp.
For constant longitudinal bunch parameters and
matched RF-voltage, higher intensity thresholds for the
 above instabilities are expected when increasing the phase slip
factor.
 
\subsection{Lowering transition energy in the SPS}

\begin{figure}[h]
\begin{center}
    \includegraphics[trim = 0 0 0 0, clip, width=0.24\textwidth]{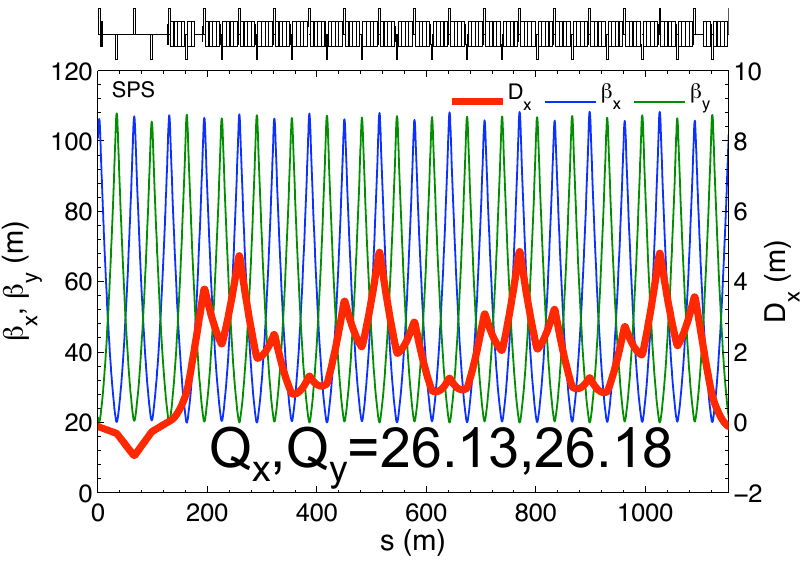}
    \includegraphics[trim = 0 0 0 0, clip, width=0.24\textwidth]{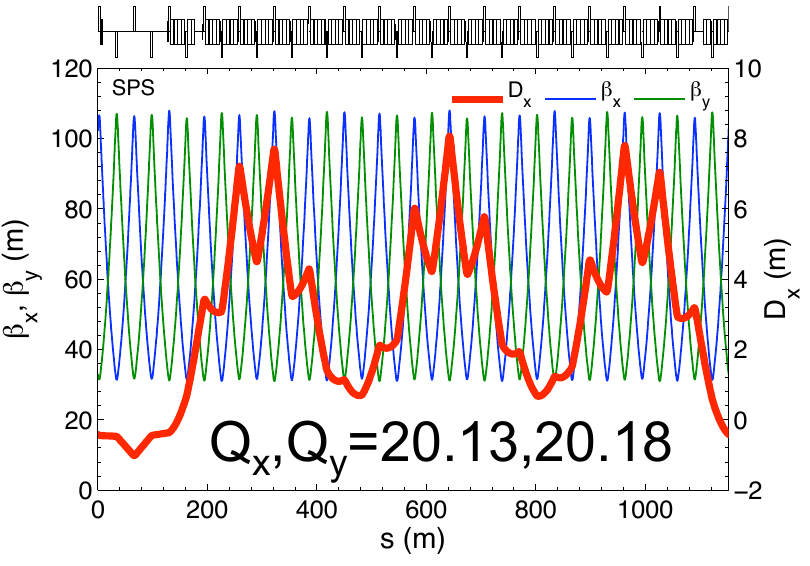}
\caption{Nominal optics (Q26) and modified (Q20) optics of the SPS (1/6 of the circumference)~\cite{OpticsConsiderations}.}
\label{FIG:NOMINALOPTICS}
\end{center}
\end{figure}

\begin{figure}[h]
    \centering
    \includegraphics[trim = 5 0 15 10, clip,width=0.6\columnwidth]{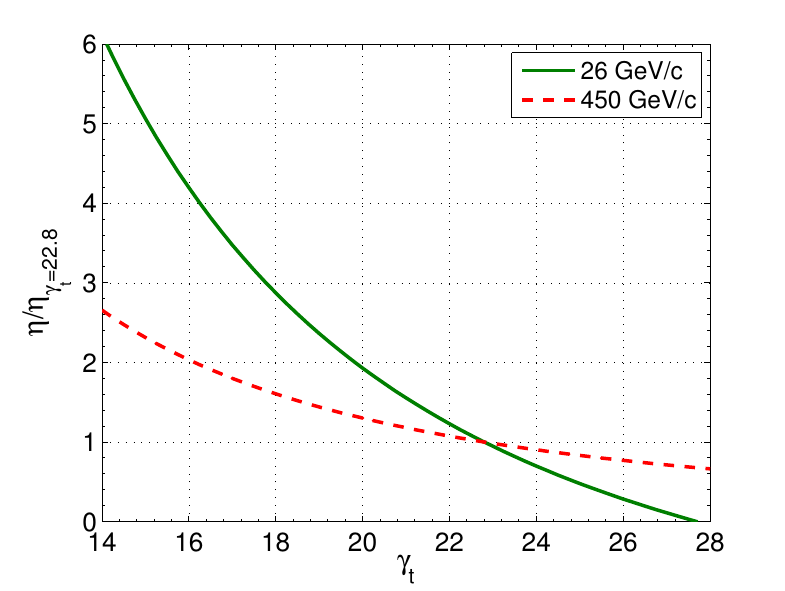}
\vspace{0pt}
    \caption{Slippage factor $\eta$ relative to the value of the nominal optics (nominal $\gamma_{t}=22.8$) as a function of $\gamma_t$~\cite{OpticsConsiderations}. 
    }
    \label{FIG:SLIPPAGEFACTOR}
    \end{figure}
The SPS has a super-symmetry of 6 with a regular FODO lattice built of 108 cells, 16 per arc and 2 per long straight section. 
In the nominal SPS optics (called Q26), the phase advance per FODO cell is close to $\pi/2$, resulting in betatron tunes between 26 and 27. Low dispersion in the long straight sections is achieved setting the arc phase advance to $4 \cdot 2\pi$. 
Figure~\ref{FIG:NOMINALOPTICS} (left) shows the optics functions in the SPS lattice for the nominal optics. 
The LHC-type proton beams  are injected at 26~GeV/c ($\gamma=27.7$), i.e. above transition ($\gamma_t=22.8$). By reducing $\gamma_t$, the slippage factor is increased throughout the acceleration cycle with the largest relative gain at injection energy, as shown in Fig.~\ref{FIG:SLIPPAGEFACTOR}, where $\eta$ normalized to the value in the nominal SPS optics ($\eta_{\mathrm{nom}}$) is plotted as a function of $\gamma_t$, for injection  and extraction energy. Significant gain of beam stability can be expected for a relatively small reduction of $\gamma_t$, especially in the low energy part of the acceleration cycle. 

In 2010, alternative optics solutions for modifying $\gamma_t$ of the SPS were investigated~\cite{OpticsConsiderations}.  
Based on the fact that in a regular FODO lattice, the transition energy is approximately equal to the horizontal tune,
$\gamma_t$ can be lowered by reducing the horizontal phase advance around the ring.
One of the possible solutions, with low dispersion in the long straight sections,
 is obtained by reducing the arc phase advance by $2\pi$,  i.e. $\mu_x,\mu_y\approx3\cdot2\pi$ and the machine tunes are close to 20 (``Q20 optics''). 
Figure~\ref{FIG:NOMINALOPTICS} (right) shows the corresponding optics functions for one super-period of the SPS. Note that in comparison to the nominal optics (``Q26''), the dispersion function follows 3 instead of 4 big oscillations along the arc with peak values increased from 4.5~m to 8~m. 
In this case, the transition energy is lowered from $\gamma_t=22.8$ in the nominal optics to $\gamma_t=18$ and $\eta$ is increased by a factor 2.85 at injection and 1.6 at extraction energy (Fig.~\ref{FIG:SLIPPAGEFACTOR}).
The maximum $\beta$-function values are the same in both optics, whereas the minima are increased by about 50\%. The optics modification is mostly affecting peak dispersion which is almost doubled. 
The fractional tunes have been chosen identical to the nominal optics in order to allow for direct comparison 
in experimental studies.

\subsection{Transverse mode coupling instability}

\begin{figure}[h]
\centering
    \includegraphics[trim = 5 0 10 0, clip, width=0.49\columnwidth]{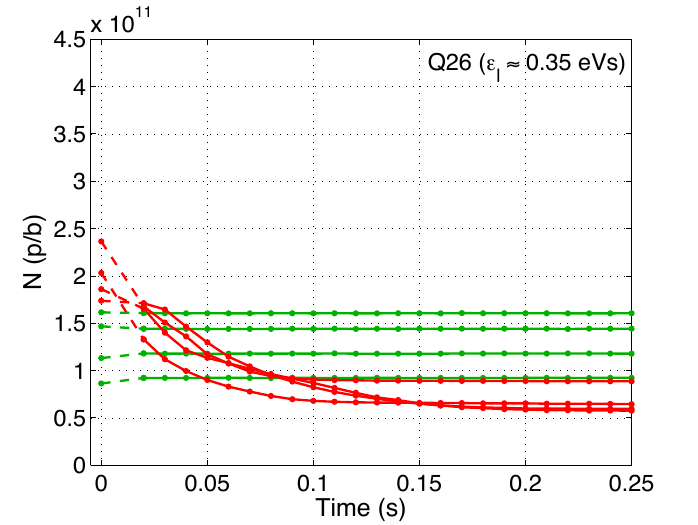}
        \includegraphics[trim = 5 0 10 0, clip, width=0.49\columnwidth]{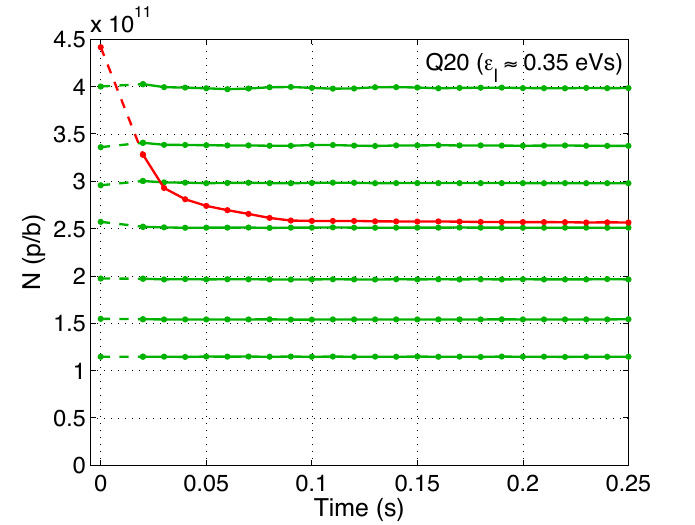}
\caption{Examples of the intensity evolution as a function of time after injection in the Q26 optics (left) and the Q20 optics (right). 
Green curves correspond to stable beam conditions, red traces indicate cases above the TMCI threshold~\cite{SPS2013}.}
\vspace{-0pt}
\label{FIG:TMCI}
\end{figure}

A series of measurements with high-intensity single bunches were conducted~\cite{Q20Experimentalstudies,IPAC2012_hannes,HB2012,SPS2013} 
in order to quantify the benefit of the Q20 optics
with respect to TMCI. In the nominal optics, the threshold is found at  1.6$\times10^{11}$~p/b, for
zero chromaticity, as shown in Fig.~\ref{FIG:TMCI} (left)~\cite{SPS2013}. In order to
pass this threshold with Q26, the vertical chromaticity has to be increased so much that the losses
are excessive due to single-particle effects. In the Q20 optics, it was demonstrated that up to 
4$\times10^{11}$ could be injected with no sign of the TMCI and low chromaticity, as shown in
Fig.~\ref{FIG:TMCI} (right)~\cite{SPS2013}. 
Such high intensity single bunches were already sent to the LHC for beam studies~\cite{Q20op}.


\subsection{Electron cloud}

\begin{figure}[h]
\centering
    \includegraphics[trim = 0 0 15 0, clip, width=0.485\columnwidth]{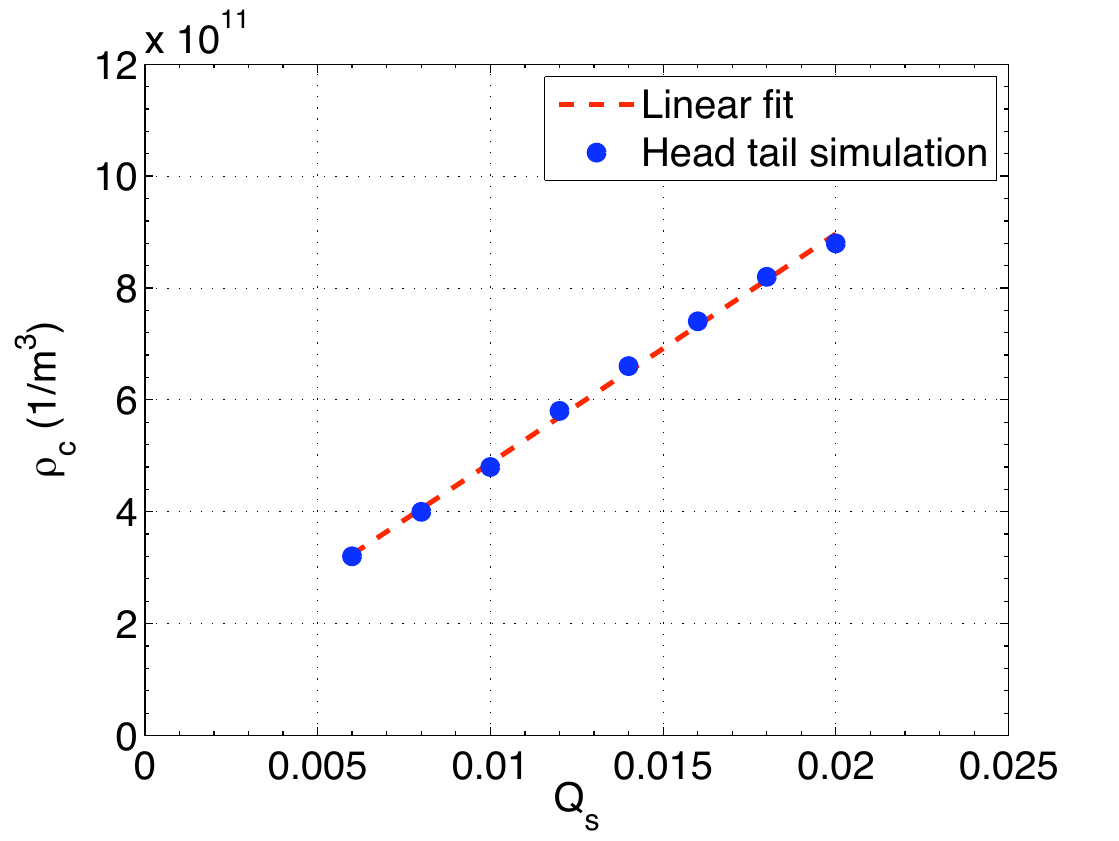}
        \includegraphics[trim = 0 0 25 0, clip, width=0.47\columnwidth]{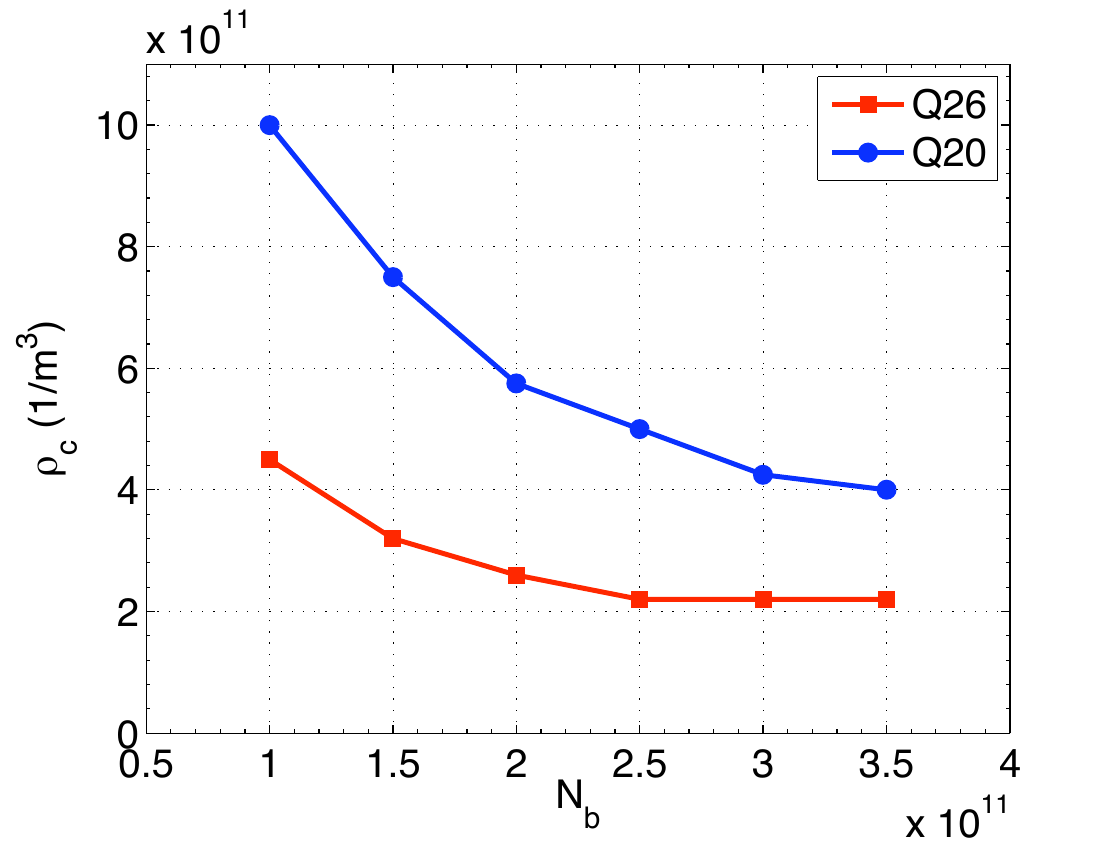}
\caption{Instability threshold density $\rho_c$ as function of the synchrotron tune 
for constant bunch parameters (left) and the predicted linear dependence. 
ECI thresholds for various intensities comparing the nominal (red) with the low $\gamma_t$ SPS optics (blue)~\cite{ecloud}.}
\vspace{-0pt}
\label{FIG:ECLOUD}
\end{figure}

Since the ECI threshold
scales with the synchrotron tune, as shown in Fig.~\ref{FIG:ECLOUD} (left)~\cite{ECImodel}, 
a clear benefit from the larger $\eta$ in the Q20 optics is expected.  Numerical
simulations were performed, assuming that the electrons are confined in bending magnets~\cite{ecloud}. 
 The expected threshold electron density $\rho_c$ for the ECI instability in the nominal (red) and the
Q20 optics (blue), as a function of the bunch intensity $N_b$ at injection energy, for matched RF voltages, is presented in Fig.~\ref{FIG:ECLOUD} (right). 
Clearly, higher thresholds are predicted for Q20.

\subsection{Longitudinal multi-bunch instabilities}
\begin{figure}[t]
\begin{center}
    \includegraphics[trim = 0 0 15 0, clip, width=0.24\textwidth]{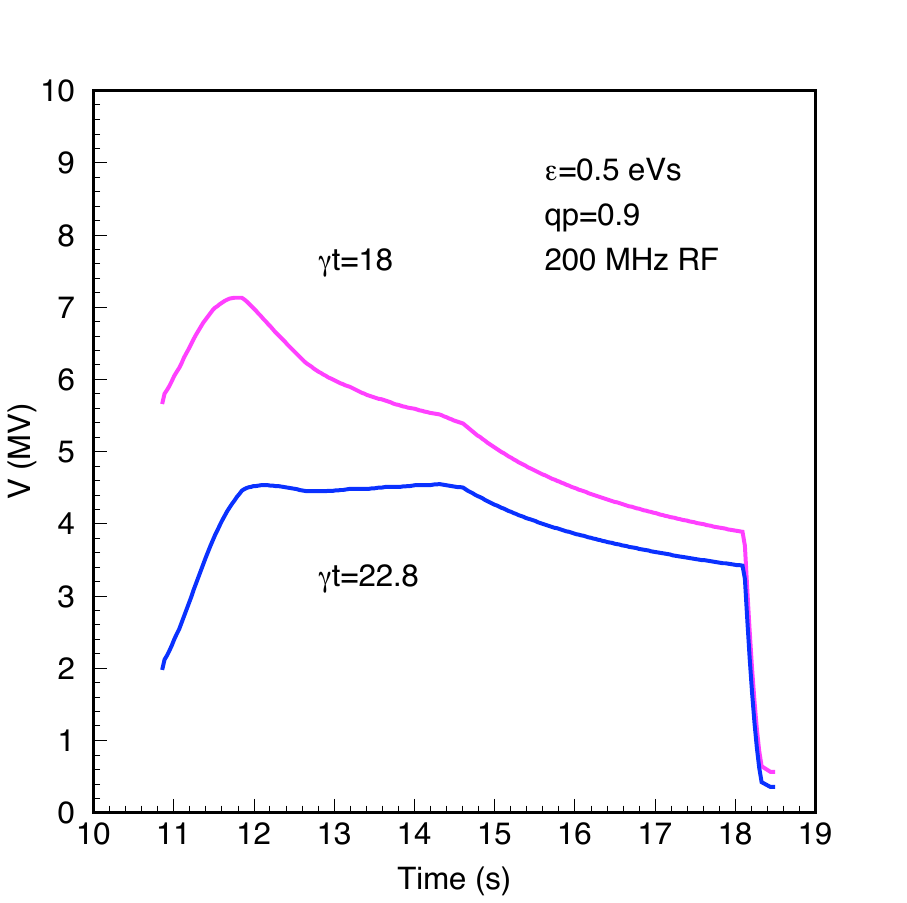}
    \includegraphics[trim = 0 0 15 20, clip, width=0.24\textwidth]{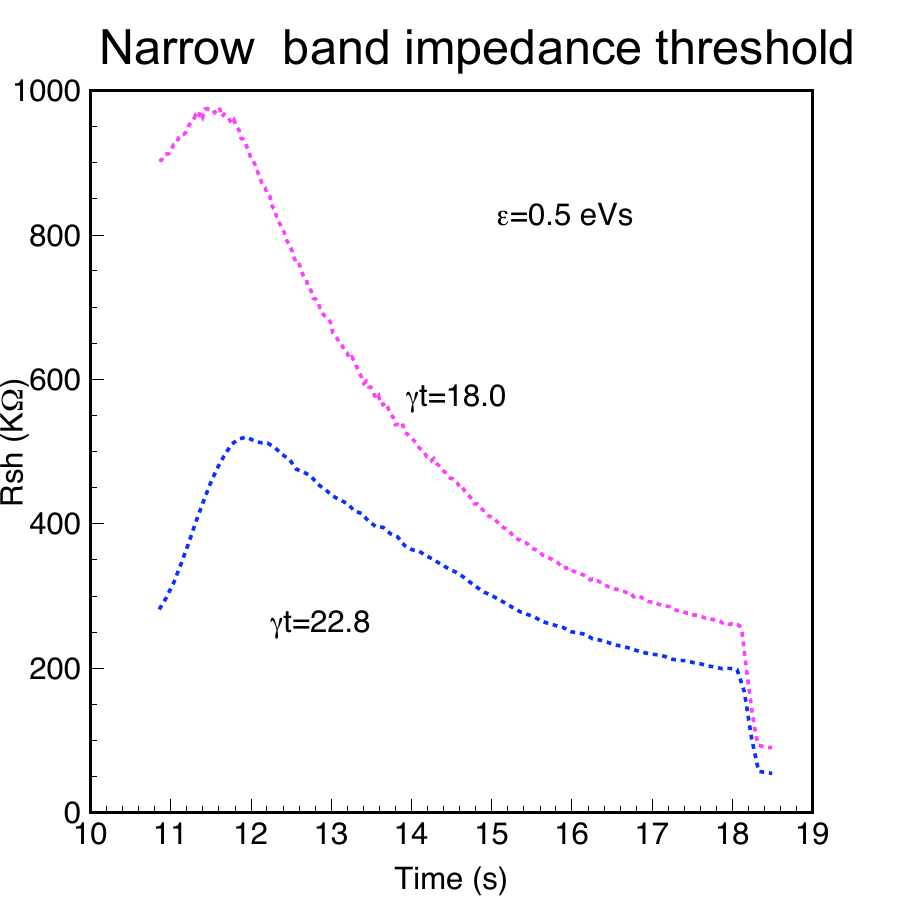}
\caption{Voltage programs (left) and narrow-band
impedance thresholds (right) through the cycle for Q26
(blue curve) and Q20 (magenta curve) optics in a single RF
system for longitudinal emittance $\epsilon_l = 0.5$~eV.s. Acceleration
starts at 10.86 s.}
\label{FIG:ThresholdComparison}
\end{center}
\end{figure}

In the Q26 nominal SPS optics the longitudinal multibunch
instability has a very low intensity threshold, which
is decreasing with the beam energy. It is expected that for
RF voltage programs providing similar beam parameters
(emittances, bunch lengths) the corresponding instability
threshold is higher in the Q20 optics. Figure~\ref{FIG:ThresholdComparison}
 presents the
calculated narrow band impedance thresholds along the cycle
for both optics in the 200~MHz single RF system for a
longitudinal emittance of $\epsilon_l = 0.5$~eV.s and the corresponding
voltage programs. For better comparison a constant
filling factor $q_p = 0.9$ (in momentum) is chosen. Note
that the impedance threshold reaches its minimal value at
flat top for both optics. 

\begin{figure}[t]
\centering
    \includegraphics[trim = 0 0 15 0, clip, width=0.49\columnwidth]{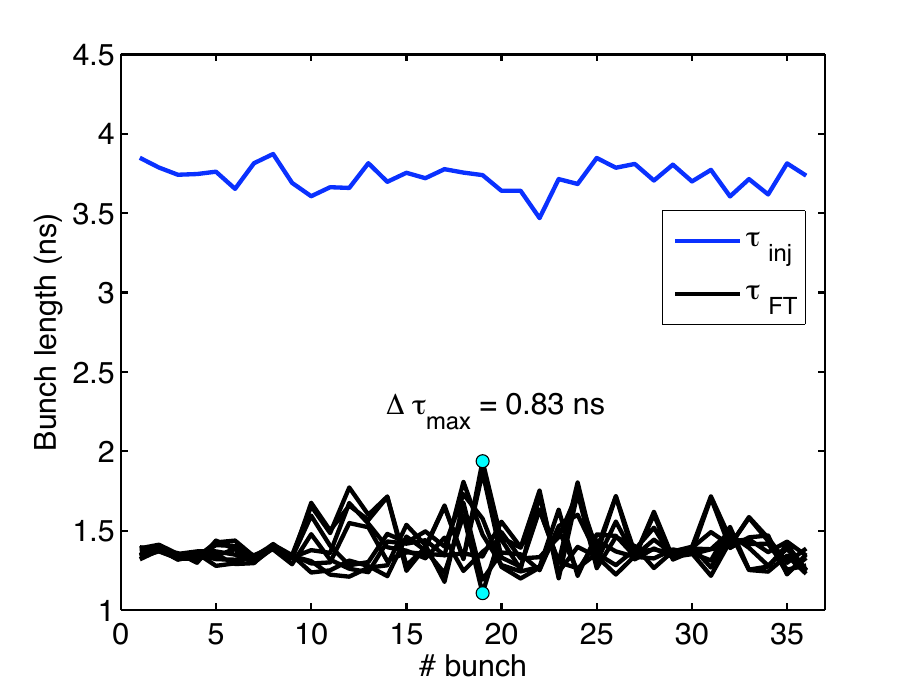}
    \includegraphics[trim = 0 0 15 0, clip, width=0.49\columnwidth]{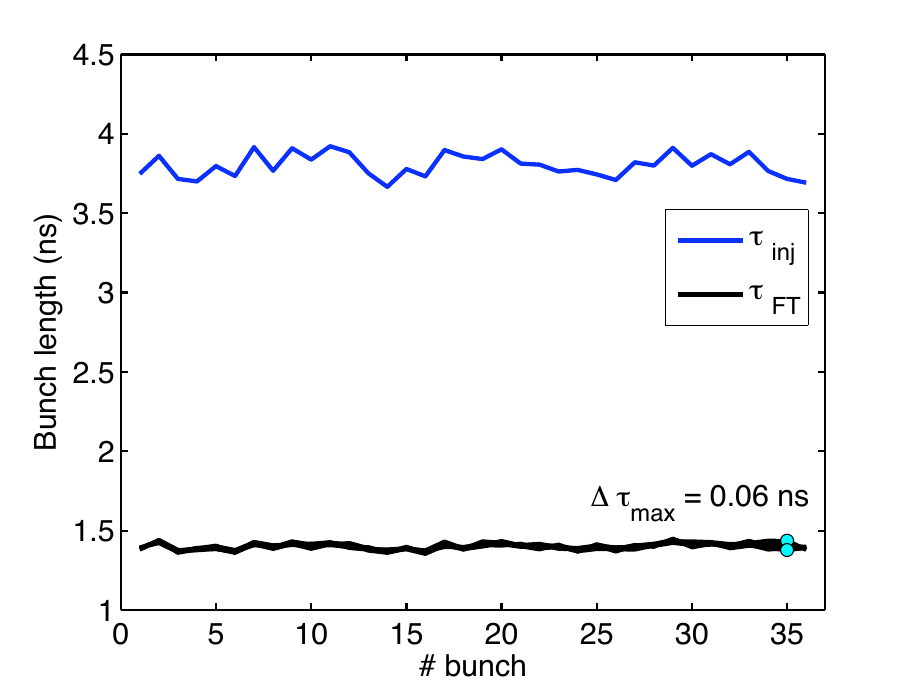}\\
    \includegraphics[trim = 0 0 15 0, clip, width=0.49\columnwidth]{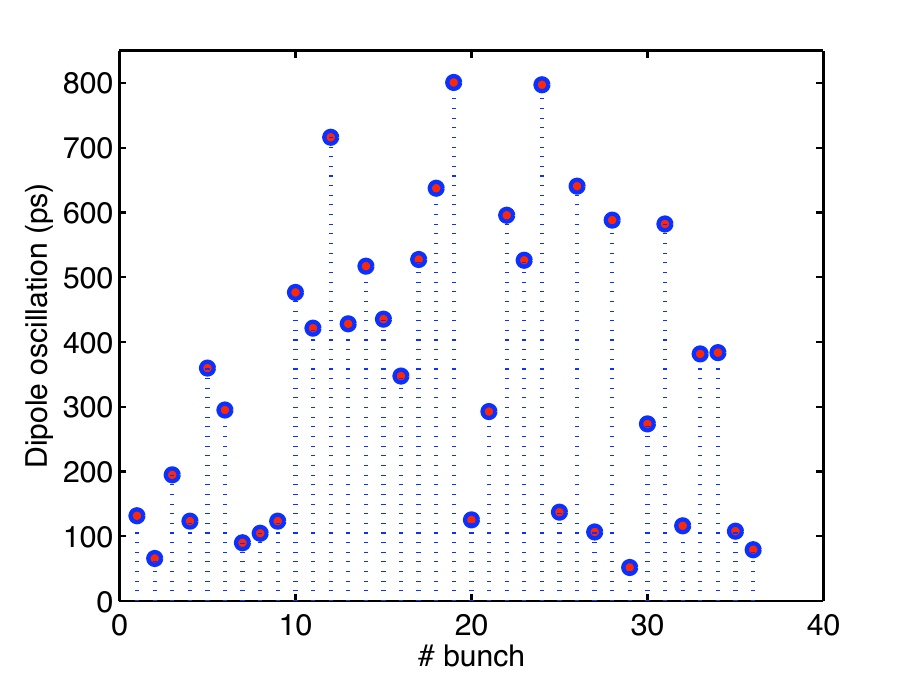}
    \includegraphics[trim = 0 0 15 0, clip, width=0.49\columnwidth]{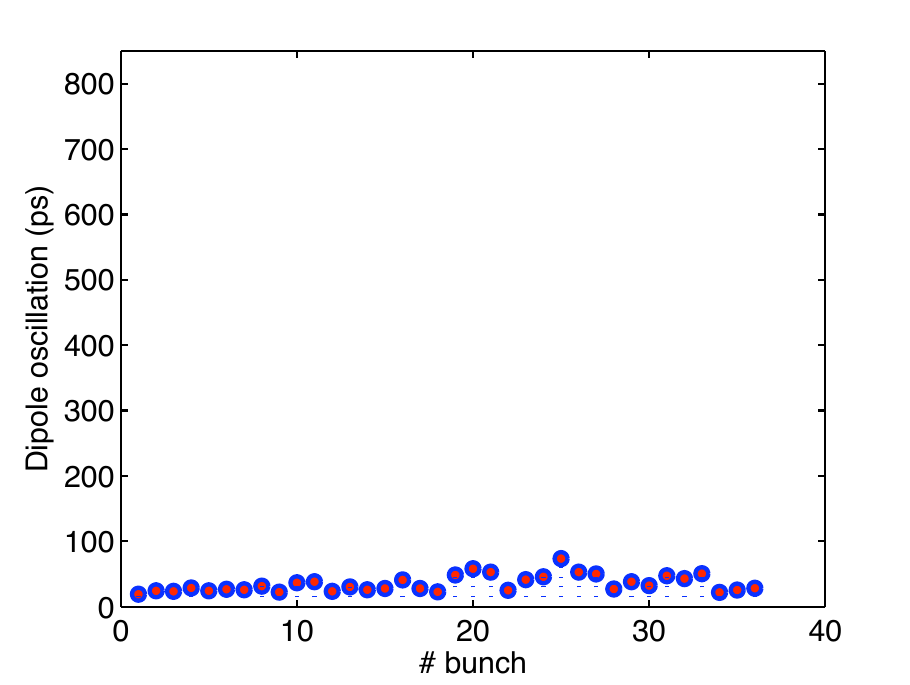}
    \vspace{-0pt}
\caption{Bunch length (top) and bunch position 
oscillations (bottom), at flat top, for the bunches of a single batch 50~ns
LHC beam, for Q26 (left) and for Q20 (right)~\cite{Q20Experimentalstudies,IPAC2012_hannes,HB2012}.}
\label{FIG:long}
\end{figure}

To stabilize the LHC beam at flat top in
the Q26 optics, controlled longitudinal emittance blow-up
is performed during the ramp, in combination with the use of a
double harmonic RF system (800~MHz) in bunch shortening
mode. The maximum voltage of the 200~MHz RF
system is needed 
in order to shorten the bunches for beam transfer to
the LHC 400~MHz bucket. Due to the limited RF voltage, 
bunches with the same longitudinal emittance at
extraction will be longer in the Q20 optics. In fact,  
for the same longitudinal bunch parameters of a stationary bucket, the required
voltage would need to be scaled with $\eta$. However, the longitudinal instability threshold at
450~GeV/c is about 50\% higher in the Q20 optics
and therefore less or no controlled longitudinal emittance
blow-up is required compared to the nominal optics,
for achieving the same beam stability. Figure~\ref{FIG:long} shows a comparison
of the beam stability (bunch length and bunch position) between the two optics, 
for one 50~ns LHC batch with $1.6\times10^{11}$ p/b.
The Q20 optics is stable even in the absence of emittance blow-up, with mean
bunch length of  around $\tau=1.45$~ns at flat top, which is compatible with injection into the LHC. 

\begin{figure}[t]
    \centering
 \includegraphics[trim = 0 4 0 0, clip, width=0.4\textwidth]{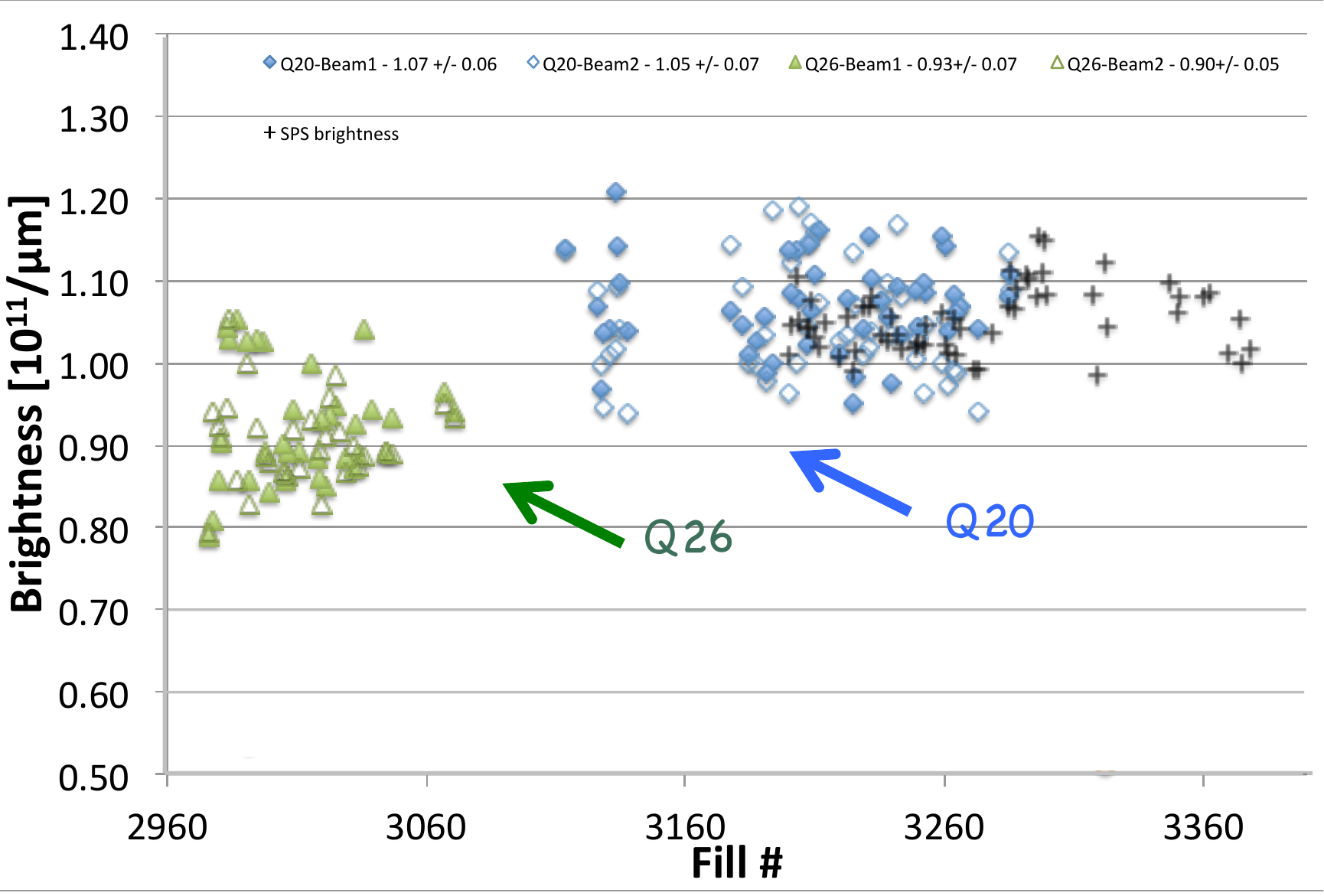}
       \caption{Average intensity over mean emittance (brightness) along the run, during the SPS operation with the nominal (green triangles)
       and the Q20 (blue diamonds) optics, in the LHC flat bottom for both beams. The SPS brightness since the Q20
       deployment is represented by the black crosses.}
        \label{LHCbright}
   \vspace{-3mm}
\end{figure}

The low transition energy optics in the SPS became operational in September 2012. 
 The switch to this new optics was very smooth, allowing very high brightness beams to be delivered to the LHC providing record luminosities~\cite{Q20op}. 
 An indication of the increased brightness delivered to the LHC is presented in Fig.~\ref{LHCbright}, where the mean bunch intensity divided by
the average of the horizontal and vertical emittance is plotted along the different LHC fills, for the second part of 2012. The green triangles represent the brightness delivered in the LHC
flat bottom using the nominal Q26 optics, whereas the blue diamonds show the one corresponding to the Q20 optics, for both beams. There was a clear brightness increase
at the LHC flat bottom of the order of 15~\% on average, due to the Q20 optics. It is also worth noting, that the SPS brightness (black crosses) is similar, demonstrating an
excellent brightness preservation between the two rings. 
This optics have been routinely used in operation during LHC run II (2015-2018) and opened the way for ultra-high brightness beams to be delivered in the HL-LHC era for protons and eventually for ions~\cite{fanouriaions}.

\section{Summary}
Using analytical and numerical methods, linear optics parameters, which have a direct impact on collective
effects, were optimised for specific examples of high-intensity, high brightness, hadron and lepton
rings. These approaches allowed a solid conceptual design of ultra-low emittance damping rings and
permitted to break intensity limitations in an existing LHC injector, without any cost impact or hardware change. 
It is certain that there is a growing need for the optics designer to transcend the 
single-particle dynamics mentality and apply such optimisation procedures for reaching the optimal performance
of rings, in design or operation.

\section{Acknowledgements}
It is a pleasure to thank G.~Arduini, T.~Argyropoulos, T.~Bohl,
H.~Braun, E.~Shaposhnikova, the LIU-SPS working group and the 
SPS operation team for invaluable contributions in several
parts of this work.


\begin{thebibliography}{99}

\bibitem{chao}
A.~W.~Chao, Physics of Collective Beam Instabilities in High Energy Accelerators, New York, Wiley, 1993.

\bibitem{theo}
 T.~Argyropoulos et al., 
 ``Loss of Landau Damping for Inductive Impedance and a Double RF System", 
 TUPWA040, IPAC13, 2013.

\bibitem{hofle}
W.~H\"ofle, 
``Progress in Transverse Feedbacks and Related Diagnostics for Hadron Machines ", 
FRXCA01, IPAC13, 2013.

\bibitem{paul}
P.~Collier et al., 
``Reducing the SPS Machine Impedance", 
EPAC02, WEPRI082, p.1458, 2002.

\bibitem{giovprl}
G.~Rumolo, et al., 
``Dependence of the Electron-Cloud Instability on the Beam Energy", 
PRL 100, 144801, 2008.

\bibitem{laslett}
L.~J.~ Laslett, 
``On Intensity Limitations Imposed by the Transverse Space- Charge Effects in Circular Accelerators", 
BNL-7534, 1963.

\bibitem{ibs}
A.~Piwinski, in Handbook of Accelerator Physics 
and Engineering, 
ed. A.~W.~Chao, M.~Tigner, World Scientific,  p.~127, 2002;
J.~Bjorken, S.~K.~Mtingwa, Part.~Accel. 13, 115, 1983.

\bibitem{fartoukh}
S.~Fartoukh, ``Achromatic telescopic squeezing scheme and application to the LHC and its luminosity upgrade", PRST-AB 16, 111002, 2013

\bibitem{transcros1}
J.~Wei, in Handbook of Accelerator Physics,
and Engineering, 
op.~cit., p.~285.
ed. A.~W.~Chao and M.~Tigner, World
Scientific, Singapore,  p.~285, 2002.

\bibitem{transcros2}
T.~Risselada, 
``Gamma transition jump schemes, 
4th CAS, CERN-91-04, p.~161, 1991.

\bibitem{Ng}
K.Y.~Ng, in Handbook of Accelerator Physics, 
and Engineering, 
op.~cit., p.~94.
ed. A.~W.~Chao and M.~Tigner, World
Scientific, Singapore,  p.~94, 2002.

\bibitem{FMC}
S.~Y.~Lee, K.~Y.~Ng, D.~Trbojevic, 
``Minimizing dispersion in flexible-momentum-compaction lattices", 
PRE, 48(4), p.~3040, 1993.

\bibitem{J-Parc}
Y.~Yamazaki ed., 
``Technical design report of J-PARC", 
KEK-2002-13, 2002.


\bibitem{PS2}
Y.~Papaphilippou et al., 
``Linear optics design of negative momentum compaction lattices for PS2", 
TH6PFP044, PAC09, p.~3805, 2009.

\bibitem{HPPS} 
Y.~Papaphilippou et al., 
``Design Options of a High-Power Proton Synchrotron for LAGUNA-LBNO'', 
THPWO081, IPAC13, 2013.

\bibitem{steve} 
S.~Hancock, 
``Gamma at Transition of the proposed PS2 Machine", 
CERN-AB-Note-2006-39, 2006.

\bibitem{bib:PS2GLASS}
H.~Bartosik et al., 
``Linear Optimization and Tunability of the PS2 Lattice", 
THPE022, IPAC10, 2010.

\bibitem{ALSGLASS}
D.~S.~Robin, W.~Wan, F.~Sannibale, and V.~P.~Suller,
``Global analysis of all linear stable settings of a storage ring lattice", PRST-AB 11, 024002, 2008.

\bibitem{DRCD}
Y.~Papaphilippou et al., 
``Conceptual design of the CLIC Damping Rings",  
TUPPC086, IPAC12, p. 1368, 2012.

\bibitem{fanouriaphd} 
F.~Antoniou, 
``Optics design of Intrabeam Scattering dominated damping rings", 
PhD Thesis, NTUA, 2013.


\bibitem{scan} 
F.~Antoniou et al., 
``Parameter scan for the CLIC Damping Rings under the influence of intrabeam scattering", 
WEPE085, IPAC10, p.~3542, 2010.

\bibitem{CIMP} 
K.~Kubo et al., 
``Intrabeam scattering formulas for high energy beams", 
PRST-AB 8, 081001, 2005.

\bibitem{newcliclat} 
H.H.~Braun et al.,
``Alternative Design of the CLIC Damping Ring Lattice", 
CLIC Note 849, 2010.

\bibitem{TMEanalytical} 
F.~Antoniou and Y.~Papaphilippou, ``Analytical considerations for linear and nonlinear optimization of
TME cells. Application to the CLIC pre-damping rings", PRST-AB 17, 064002, 2014.


\bibitem{PhysRevAccelBeams.22.091601}
  S.~Papadopoulou, F.~Antoniou, and Y.~Papaphilippou, ``Emittance reduction with variable bending magnet strengths: Analytical optics considerations and application to the Compact Linear Collider damping ring design", PRAB 22, 091601, 2019.


\bibitem{wiggler}
D.~Schoerling et al., ``Design and system integration of the superconducting wiggler magnets for the Compact Linear Collider damping rings", PRST-AB 15, 042401, 2012.

\bibitem{Fajardo:2016nnq}
L.~G.~Fajardo, F.~Antoniou, A.~Bernhard, P.~Ferracin, J.~Mazet, S.~Papadopoulou, Y.~Papaphilippou, J.~Pérez and D.~Schoerling,
``Design of Nb3Sn Wiggler Magnets for the Compact Linear Collider and Manufacturing of a Five-Coil Prototype,''
IEEE Trans. Appl. Supercond.,26 no.4, 4100506, 2016. 

\bibitem{elena}
LIU-SPS Beam Dynamics Working Group, chaired by E.~Shaposhnikova,  http://paf-spsu.web.cern.ch/paf-spsu/

\bibitem{OpticsConsiderations}
H.~Bartosik G. Arduini, Y. Papaphilippou, 
``Optics considerations for lowering transition energy in the SPS'', 
MOPS012, IPAC11, p.~619, 2011.
%

\bibitem{Q20Experimentalstudies}
H.~Bartosik et al., 
``Experimental studies with low transition energy optics in the SPS''
MOPS010, IPAC11, p.~613, 2011.

\bibitem{IPAC2012_hannes}
H.~Bartosik et al., ``Increasing Instability Thresholds in the SPS by Lowering Transition Energy'',
 WEPPR072, IPAC12, p.~3096, 2012; 
 
 \bibitem{HB2012}
 H.~Bartosik et al.,
 ``Low-gamma transition optics for the SPS: simulation and experimental results for high brightness beams",
WEO1B01, Proceedings of HB2012, Beijing, China, 2012.
	
\bibitem{SPS2013}
H.~Bartosik et al., 
``Experimental studies for future LHC beams in the SPS", 
TUPME034, IPAC13, 2013.

\bibitem{Q20op} 
Y.~Papaphilippou et al., 
``Operational Performance of the LHC Proton Beams with the SPS Low Transition Energy Optics"
THPWO080, IPAC13, 2013.


\bibitem{ECImodel}
K. Ohmi and F. Zimmermann, 
``Head-Tail Instability Caused by Electron Clouds in Positron Storage Rings'', 
PRL 85, 3831, 2000.

\bibitem{ecloud}
H.~Bartosik et al., ``Impact of low transition energy optics to the electron cloud instability of LHC beams in the SPS", 
MOPS011, IPAC11, p.~616, 2011.

\bibitem{fanouriaions}
F.~Antoniou et al., 
``Considerations on SPS low-transition energy optics for LHC ion beams", 
TUPME046, IPAC13, 2013.














\end{thebibliography}
\end{document}